\documentclass[reprint,showpacs,amsmath,amssymb,aps]{revtex4-1}
\usepackage{dcolumn}% Align table columns on decimal point
\usepackage{bm}% bold math
\usepackage{mathrsfs}
\usepackage[titletoc]{appendix}
\usepackage{graphicx}
\usepackage{float}
\usepackage{tikz}

\begin{document}

\title{Anomalous Geometric Spin Hall Effect of Light?}
\author{Zhen-Lai Wang}
\author{Xiang-Song Chen}%
\email{cxs@hust.edu.cn}
\affiliation{School of Physics and
MOE Key Laboratory of Fundamental Quantities Measurement,
Huazhong University of Science and Technology, Wuhan 430074, China}

\date{\today}
\begin{abstract}
The geometric spin Hall effect of light (GSHEL), similar to the spin Hall effect of light, is also a spin-dependent shift of the centroid of light beam's intensity (energy flux), but it is a purely geometric effect that does not depend on a particular light-matter interaction. In this paper, we discuss the GSHEL with respect to momentum instead of energy flux, and find out  that in the case of the symmetric energy-momentum tensor,  the shift of the centroid of momentum flux is  double  that of energy flux. Interestingly,  for the canonical energy-momentum tensor,  the centroid shift of momentum flux  agrees with that of energy flux. If we consider the effect of orbital angular momentum, however, the centroid displacement of momentum flux is twice that of energy flux for both energy-momentum tensors. To tell which energy-momentum tensor of light field would be more ``correct'', we propose a  experimental scheme to test the GSHEL of momentum flux through the mechanical effect of light. 
\end{abstract}


\maketitle

\emph{Introduction and key note}---
It is well known that light field can carry  energy, momentum and angular momentum, which play important roles in the light-matter interaction.  Nowadays, many techniques have been realized to manipulate microscopic particles by using the linear and angular momentum of light, such as laser cooling, optical tweezers and optical spanners \cite{Simp96,Grier03}.  As reciprocal influences of the matter upon the light, the  spin Hall effect of light (SHEL) has  attracted  a considerable amount of theoretical and experimental investigations. It is a novel phenomenon predicting a spin-dependent shift of the centroid of light beam's intensity.  In fact, there are two polarization-dependent shifts of a  light beam when it was reflected or refracted from an optical interface, namely  the longitudinal Goos-H\"anchen shift and the transverse Fedorov-Imbert shift \cite{Onod04,Blio06,Host08,Blio15}  (for a reference, see \cite{Blio13}). The latter is regarded as an example of the  SHEL, which now is usually interpreted as the spin-orbital interaction of light in terms of the  Berry phase \cite{Berr84}.

In 2009, another type of SHEL, named geometric SHEL (GSHEL), was  proposed \cite{Aiel09}. This effect says that a spin-dependent transverse displacement of the light intensity centroid is observed  in a plane not perpendicular to the propagation of the light beam. It originates from the nonzero transverse angular momentum observed in the detector frame.  Unlike the conventional  SHEL that requires the light-matter interaction, the GSHEL  is  of purely geometric nature. Later on, the orbital angular momentum of light beam was shown to cause a transverse shift in addition to the shift caused by spin \cite{kong12}. In 2014,  it was reported that the spatial intensity centroid of a  polarized light beam transmitted across an oblique polarizer \cite{korg14} underwent a displacement larger than the conventional  SHEL, which was claimed as the observation of the GSHEL. So far, the GSHEL or similar effects have been analyzed for collimated paraxial beams \cite{Beks09}, tightly focused vector beams \cite{Neug14} and inhomogeneous polarized beam \cite{Ling17}. For the convenience of reference and comparison,  we put here the centroid displacement of GSHEL with respect to energy flux
\begin{equation}\label{Y}
{\langle y \rangle}_P=\int y~\overline{T}^{z0}_{\rm sym}\mathrm{d}x\mathrm{d}y{\Big /}\int \overline{T}^{z0}_{\rm sym}\mathrm{d}x\mathrm{d}y\simeq\frac{\lambda}{4\pi}\sigma\tan{\theta}.
\end{equation}
 ${\langle y \rangle}_P$ denotes the shift of the barycenter and the subscript $P$ indicates that the barycenter is evaluated with respect to the Poynting vector ( energy flux $T^{i0}_{\rm sym}$). The bar over $T^{z0}_{\rm sym}$ denotes time-averaging.  $\sigma=\pm1$ is  the polarization of light and $\lambda$ the wavelength. $\theta$ is the tilted angle between the detector plane and the transverse plane of the light beam. FIG. 1 is a schematic diagram of the beam and detection system. Strictly speaking,  Eq. ({\ref{Y}}) receives correction for the angular spread of the beam, which we omit in the following discussion. 
 
 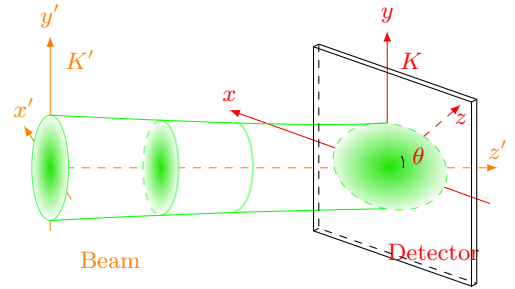
\begin{figure}[H]\label{fig}
\centering
\begin{tikzpicture}[scale=0.7]
	\draw[line width=0.4pt,-latex,orange] node [right=0.4cm,above=1.2cm] {$K'$}  (0,-1.2)--(0,2.5) [above] node{$y'$};
	\draw[line width=0.4pt,-latex,orange]  node [right=0.8cm,below=1cm]{Beam} (0.4,-0.6)--(-0.5,0.8) [above] node{$x'$};
	\draw[line width=0.4pt,-latex,dashed,orange](0,0)--(8.5,0)[above] node{$z'$};
	  \draw[line width=0.4pt,-latex,dashed,red](6.4,0)--(7.8,1.2)[below] node{$z$};
	     \draw[line width=0.4pt,-latex,red] node [right=4.8cm,above=1.2cm] {$K$}  (6.4,0) -- (6.4,2.6) [above] node{$y$};
   \draw[line width=0.4pt,-latex,red](8.35,-0.68)--(3.4,1.1) [above] node{$x$}; 
	 \shadedraw[inner color=green!90!red,outer color=white,draw=green,opacity=0.7] (0,0) ellipse [x radius=0.35cm,y radius=1.0cm];	 
	 \draw[inner color=green!90!red,outer color=white,draw=white,dashed,opacity=0.7] (2.1,0) ellipse [x radius=0.34cm,y radius=0.9cm];
	 	 \draw[line width=0.8pt,draw=green,dashed,opacity=0.4] (2.1,0.90) arc  (90:270:0.34 and 0.9) ;	
	 \draw[line width=0.8pt,draw=green,opacity=0.4] (2.1,-0.90) arc  (-90:90:0.34 and 0.9) ;	
	 	  \draw[line width=0.8pt,draw=green,opacity=0.4] (3.5,-0.82) arc  (-90:90:0.35 and 0.84) ;
	 	  \shadedraw[inner color=green!90!red,outer color=white,draw=green,dashed,opacity=0.7,rotate=-15.9] (7.3,1.79)  arc (0:360:1.1 and 0.8);
	 	 \draw[color=green] (0,1) .. controls (4,0.8) and (5.5,0.8) .. (6.4,0.85);
	 	  	 \draw[color=green] (0,-1) .. controls (4,-0.8) and (5.5,-0.8) .. (6.35,-0.78);
	
   	\draw (5,2.3) -- (5,-1.2) -- (8,-2.25)--(8,1.25) -- (5,2.3) node [right=1.6cm,below=2.5cm,red]{Detector};
   		\draw[dashed](5.1,2.35) -- (5.1,-1.15) -- (8.1,-2.2);      		 	
   	\draw (8.1,-2.2)--(8.1,1.3)-- (5.1,2.35);
   	\draw (5,2.3)--(5.1,2.35) (5,-1.2)--(5.1,-1.15) (8.0,-2.25)--(8.1,-2.2)(8.0,1.25)--(8.1,1.3);
   	\draw (6.7,0) arc [start angle=-15, end angle=30, radius=0.3cm,red] node [right=0.03cm,red]{$\theta$};
	\end{tikzpicture}
\caption{The light beam propagates along the $z^{\prime}$ axis, around which the beam is rotationally symmetric. The $x-y$  plane of the laboratory frame $K$  is in the detection plane. The $y$  axis is parallel to the $y^{\prime}$  axis of the beam frame $K^{\prime}$. The  $z$ axis is tilted by an angle $\theta$  with respect to the $z^{\prime}$ axis. If the light beam carries angular momentum along the $z^{\prime}$ direction, a transverse angular momentum $J^{x}$ can be  observed in the $K$ frame.}
\end{figure}
 
Light carries energy and its energy transportation is represented by energy flux. Likewise, light also carries momentum and we use the momentum flux to describe  momentum transportation. In fact, the energy-momentum (E-M) tensor of light field, which bands together the energy density, momentum density, energy flux density and momentum flux density, is a very convenient tool to analyze the properties of light. In this paper, we discuss the GSHEL with respect to the momentum flux density, and reach a result different from Eq. (\ref{Y}). To show the difference, for a beam with only spin polarization we get the shift of the barycenter of the momentum flux $T^{zz}_{\rm sym}$: 
\begin{equation}\label{Ys}
{\langle y_{\rm sym} \rangle}_T=\int y~\overline{T}^{zz}_{\rm sym}\mathrm{d}x\mathrm{d}y{\Big /}\int \overline{T}^{zz}_{\rm sym}\mathrm{d}x\mathrm{d}y=\frac{\lambda}{2\pi}\sigma\tan{\theta}.
\end{equation}
This result differs from the conventional GSHEL with respect to  energy flux by a factor of 2, so we call it anomalous GSHEL. $T^{z0}_{\rm sym}$ in Eq. (\ref{Y}) and $T^{zz}_{\rm sym}$ in Eq. (\ref{Ys}) corresponds to $z$-direction  energy flux density and $z$-direction flux density of $z$-direction momentum of the symmetric E-M tensor:
 \begin{equation}
T^{\mu\nu}_{\rm sym}=-F^{\mu \alpha}F^{\nu}_{~\alpha}+\frac{1}{4}g^{\mu\nu}F^{\alpha\beta}F_{\alpha\beta},
\end{equation}
where $g^{\mu\nu}$ is the  Minkowski metric with signature $(+---)$. 

As is well-known, the expression of E-M tensor is not uniquely determined by the energy and momentum conservation laws. Besides the symmetric E-M tensor, we also have the well-known canonical E-M tensor:
 \begin{equation}\label{Tcan}
T^{\mu\nu}_{\rm can}=-F^{\mu \alpha}\partial^{\nu}A_{\alpha}+\frac{1}{4}g^{\mu\nu}F^{\alpha\beta}F_{\alpha\beta}.
\end{equation}
If we calculate the barycenter of the momentum flux according to the canonical E-M tensor, we get the result
\begin{equation}\label{Yc}
{\langle y_{\rm can} \rangle}_T=\int y\overline{T}^{zz}_{\rm can}\mathrm{d}x\mathrm{d}y{\Big /}\int\overline{T}^{zz}_{\rm can}\mathrm{d}x\mathrm{d}y=\frac{\lambda}{4\pi}\sigma\tan{\theta}.
\end{equation}
The prediction  of  the canonical version in Eq. (\ref{Yc}) coincides with Eq. (\ref{Y}), but that of  the symmetric version in Eq. (\ref{Ys}) does not. For this reason, we argue that the GSHEL can be used as a strong criteria to single out a more valid version.  In other words, either or both of them must be wrong. In the next section, we will derive the above results in a simple way. Then we will discuss the experimental scheme to test the GSHEL with respect to momentum flux.
 
\emph{GSHEL or Anomalous GSHEL}---
We know the GSHEL is related to the nonzero transverse angular momentum of light beam. In this section, we will deduce the above results by using the sum rule of angular momentum. 

The beam frame and laboratory frame are connected by a rotation transformation, i.e. $x^{\mu}=\Lambda^{\mu}_{~\nu}x^{\prime\nu}$ [or $ x^{\prime\mu}=(\Lambda^{-1})^{\mu}_{~\nu}x^{\nu}$]. Here the rotation transformation matrix is 
\begin{equation}
\Lambda^{\mu}_{~\nu}=\begin{bmatrix}
      1&0&0&0\\
     0&\cos\theta&0&\sin\theta \\
     0&0&1&0\\
      0&-\sin\theta&0&\cos\theta \\
\end{bmatrix}.
\end{equation}
Then the momentum flux in the detection frame relates to that of the beam frame by
\begin{eqnarray}
T^{zz}(x)&=&\Lambda^{z}_{~\alpha}\Lambda^{z}_{~\beta}T^{\prime\alpha\beta}(x^{\prime})=-(T^{\prime xz}+T^{\prime zx})\sin \theta \cos \theta\nonumber\\
&~&~~~~+( T^{\prime zz}\cos ^2 \theta+T^{\prime xx}\sin ^2 \theta).
\end{eqnarray}
In the beam frame, according to the axial symmetry of the beam around its beam axis,  $T^{\prime zz}$ should be even function of the coordinates $x^{\prime}$  and $y^{\prime}$. Furthermore, $T^{\prime xx}$ can be ignored compared to  $T^{\prime zz}$ because the light beam mainly carries momentum along the propagation direction and the momentum is mainly transported in the propagation direction. Hence, we obtain 
\begin{eqnarray}\label{Yt}
&~&{\langle y \rangle}_T=\int y~\overline{T}^{zz}\mathrm{d}x\mathrm{d}y{\Big /}\int ~\overline{T}^{zz}\mathrm{d}x\mathrm{d}y\nonumber\\
&~&=-\tan\theta\int y~(\overline{T}^{\prime xz}+\overline{T}^{\prime zx})\mathrm{d}x\mathrm{d}y{\Big /}\int ~\overline{T}^{\prime zz}\mathrm{d}x\mathrm{d}y\nonumber\\
&~&=-\tan\theta\int y^{\prime}(\overline{T}^{\prime xz}+\overline{T}^{\prime zx})\mathrm{d}x^{\prime}\mathrm{d}y^{\prime}{\Big /}\int \overline{T}^{\prime zz}\mathrm{d}x^{\prime}\mathrm{d}y^{\prime}.
\end{eqnarray}
In the last step we have transformed the area element of the laboratory frame $K$ to that of the beam frame $K^{\prime}$ without changing the final result.

To proceed with  the expression in Eq. (\ref{Yt}), we have two E-M tensors in hand, the canonical one and the symmetric one. 
\begin{equation}
{\langle y_{\rm sym} \rangle}_T=-\dfrac{2\displaystyle\int y^{\prime}~\overline{T}^{\prime zx}_{\rm sym}\mathrm{d}x^{\prime}\mathrm{d}y^{\prime}}{\displaystyle\int ~\overline{T}^{\prime zz}_{\rm sym}\mathrm{d}x^{\prime}\mathrm{d}y^{\prime}}\tan\theta,
\end{equation}
\begin{equation}
{\langle y_{\rm can} \rangle}_T=-\dfrac{\displaystyle\int y^{\prime}(\overline{T}^{\prime xz}_{\rm can}+\overline{T}^{\prime zx}_{\rm can})\mathrm{d}x^{\prime}\mathrm{d}y^{\prime}}{\displaystyle\int ~\overline{T}^{\prime zz}_{\rm can}\mathrm{d}x^{\prime}\mathrm{d}y^{\prime}}\tan\theta.
\end{equation}
According to the axial symmetry of the light beam, we have 
\begin{eqnarray}
&~&\int -y^{\prime}~\overline{T}^{\prime zx}\mathrm{d}x^{\prime}\mathrm{d}y^{\prime}=\int x^{\prime}~\overline{T}^{\prime zy}\mathrm{d}x^{\prime}\mathrm{d}y^{\prime}\nonumber\\
&~&~~~~~=\frac{1}{2}\int (x^{\prime}~\overline{T}^{\prime zy}-y^{\prime}~\overline{T}^{\prime zx})\mathrm{d}x^{\prime}\mathrm{d}y^{\prime}
\end{eqnarray}
and
\begin{equation}
\int ~\overline{T}^{\prime zz}_{\rm sym}\mathrm{d}x^{\prime}\mathrm{d}y^{\prime}\simeq\int ~\overline{T}^{\prime zz}_{\rm can}\mathrm{d}x^{\prime}\mathrm{d}y^{\prime}\simeq P_{z}^{\prime}=n\hbar k,
\end{equation}
where $n$ is the photon number per unit time cross the plane $x^{\prime}-y^{\prime}$, namely the photon number flux.

For the symmetric E-M tensor, $M^{\prime zxy}_{\rm sym}=x^{\prime}\,T^{\prime zy}_{\rm sym}-y^{\prime}\,T^{\prime zx}_{\rm sym}$ represents the flux of total angular momentum along the direction of propagation. Hence, for a beam with only spin polarization we obtain
 \begin{eqnarray}
{\langle y_{\rm sym} \rangle}_T&=&\tan\theta\int (x^{\prime}~\overline{T}^{\prime zy}_{\rm sym} -y^{\prime}~\overline{T}^{\prime zx}_{\rm sym})\mathrm{d}x^{\prime}\mathrm{d}y^{\prime}{\Big /}P_{z}^{\prime}\nonumber\\
&=&\dfrac{n\sigma\hbar}{n \hbar k}\tan\theta=\frac{\lambda}{2\pi}\sigma\tan{\theta}.
\end{eqnarray}

For the canonical E-M tensor,  $L^{\prime zxy}_{\rm can}=x^{\prime}\,T^{\prime zy}_{\rm can}-y^{\prime}\,T^{\prime zx}_{\rm can}$  gives merely the flux of orbital angular momentum along the propagation direction. Thus, for a beam with only spin polarization, we have $\displaystyle\int y\, \overline{T}^{\prime zx}_{\rm can}\,\mathrm{d}x\mathrm{d}y=0$. For the collimated light beam, the light wave function is approximately in the simultaneous eigenstate of energy and longitudinal momentum. Thus, from the expression of the canonical E-M tensor, we can observe the following relation
\begin{equation}
 \overline{T}^{\prime xz}_{\rm can}\simeq \frac{k}{\omega} \overline{T}^{\prime x0}_{\rm can}=c\, \overline{T}^{\prime x0}_{\rm can}.
\end{equation}
Then we have
\begin{equation}\label{Yc2} 
{\langle y_{\rm can} \rangle}_T\simeq -c\,\tan\theta\int y^{\prime}~\overline{T}^{\prime x0}_{\rm can}\,\mathrm{d}x^{\prime}\mathrm{d}y^{\prime}{\Big /}P_z^{\prime}.
\end{equation}
Again, due to the axial symmetry of the light beam, we arrive at 
\begin{eqnarray}\label{J}
&~&\int -y^{\prime}~\overline{T}^{\prime x0}_{\rm can}\,\mathrm{d}x^{\prime}\mathrm{d}y^{\prime}=\int x^{\prime}~\overline{T}^{\prime y0}_{\rm can}\,\mathrm{d}x^{\prime}\mathrm{d}y^{\prime}\nonumber\\
&~&~~~~~~~=\frac{1}{2}\int (x^{\prime}~\overline{T}^{\prime y0}_{\rm can}-y^{\prime}~\overline{T}^{\prime x0}_{\rm can})\,\mathrm{d}x^{\prime}\mathrm{d}y^{\prime}.
\end{eqnarray}
Note importantly that the Poynting vector $({\bm E}\times {\bm B})^{i}$ is both the energy flux density $T^{i0}_{\rm sym}$ and momentum density $T^{0i}_{\rm sym}$ of the symmetric E-M tensor, and it is also the energy flux density $T^{i0}_{\rm can}$ of the canonical E-M tensor in the radiation gauge \cite{Van,Chen,Lead16,Lead14}. Hence, Eq. (\ref{J}) represents the total time-averaged angular momentum  of the beam per unit length and the final result of Eq. (\ref{Yc2}) is
 \begin{equation}
{\langle y_{\rm can} \rangle}_T=\dfrac{cN_s\sigma\hbar}{2n \hbar k}\tan\theta=\frac{\lambda}{4\pi}\sigma\tan{\theta}.
\end{equation}
Here $N_s$ is the photon number per unit length along the direction of propagation and we have $n=cN_s$.

So far we have proved the main results presented in the first section. If the light beam carries orbital angular momentum $l \hbar$  as well as spin angular momentum $\sigma\hbar$ per photon along the direction of propagation, we can easily repeat the above analysis and have the following more general results about the GSHEL:
\begin{eqnarray}
{\langle y_{\rm sym} \rangle}_P&=&{\langle y_{\rm can} \rangle}_P=\frac{\lambda}{4\pi}(l+\sigma)\tan{\theta}\label{ypsl},\\
{\langle y_{\rm sym} \rangle}_T&=&\frac{\lambda}{4\pi}(2l+2\sigma)\tan{\theta}\label{ytslc},\\
{\langle y_{\rm can} \rangle}_T&=&\frac{\lambda}{4\pi}(2l+\sigma)\tan{\theta}\label{ytsls}.
\end{eqnarray}
It is valuable to make two remarks on the above results:

(i) As we have seen in Eq.(\ref{Yt}), there are two pieces of momentum flux: The first piece is the transverse flow of the longitudinal momentum $T^{\prime xz}$, which is  approximately proportional to the transverse component flow of the energy both for the symmetric and canonical E-M tensors and its rotation around the direction of the propagation gives the total angular momentum flux; the second piece is the longitudinal flow of the transverse momentum $T^{\prime zx}$, and for the symmetric E-M tensor its rotation refers to the total angular momentum flux, but for the canonical  E-M tensor its rotation only refers to the orbital angular momentum flux.

(ii) Therefore, there are two kinds of anomalous GSHEL: One represents the comparison  between  GSHEL of momentum flux and that of energy flux, and the difference originates from the two pieces of momentum flux. Another refers to the comparison  between  GSHEL of orbital angular momentum and that of spin, or in other words, the different predictions of  the symmetric and canonical E-M tensors for the spin-generated GSHEL.  

Thus,  the GSHEL of orbital angular momentum cannot be used to discriminate these two E-M tensors, since in Eq. (\ref{ypsl}-\ref{ytsls}), they both give the same centroid shifts for both cases of energy flux and momentum flux. However, from Eq. (\ref{ytslc}) and (\ref{ytsls}), we could conclude that the GSHEL with respect to momentum flux of spin-polarized beam can serve as a probe for the possible experimental test of those two forms of E-M tensor. It should be stressed that, although these results is interpreted in terms of photon, they could  be also derived with wave packets in classical electromagnetism.

\emph{GSHEL Manifested as the Moment of Force}---
In the preceding section,  we presented the GSHEL in terms of  the momentum flux  and angular momentum flux. We know that the momentum and angular momentum of light field can manifest as mechanical effect by interaction with matter; namely, light field is capable of exerting force and torque on the matter. According to the above conclusion, the key difference of the predictions based on different E-M tensors is originated from spin angular momentum, so we should use the spin-polarized light beam to perform the test.  Naturally, we consider the torque effect of the transverse angular momentum of light beam. For the spin-polarized light beam carrying orbital angular momentum, we can easily get 
\begin{equation}\label{Tor}
\tau^{x}=\left\langle\dfrac{{\rm d }J^{x}}{{\rm d}t} \right\rangle\simeq -\int \overline{M}^{zyz}{\rm d}x {\rm d}y =n(\sigma+l)\hbar\sin\theta,
\end{equation}
which means the light beam exerts a torque to the detector around the $x$ direction. Eq. (\ref{Tor}) is sound  for both the canonical E-M tensor and the symmetric one. It implies that the global torque effect fails  to distinguish between the symmetric E-M tensor and the canonical one.

\begin{figure}[H]
\centering
\begin{tikzpicture}[fill opacity=0.8,scale=0.4]  
\fill[draw=white, fill=black] (-0.3,-0.3) rectangle (8.3,8.3);
\draw [draw=white, fill=black] (0,0) grid  (8,8) rectangle (0,0);
\draw[orange,thick,-latex] (-2.5,4) -- (11.5,4) node[right] {$x$};
\draw[orange,thick,-latex] (4,-1) node[red,below=5pt] {Detector}-- (4,9.5) node[above] {$y$};
\fill[draw=red, fill=black] (6.01,7.01) rectangle (6.99,7.99);
\fill[draw=red, fill=black] (6.01,0.01) rectangle (6.99,0.99);
\draw[red,-latex] (6.5,8) -- (7.2,9.2) node[right] {Detection Element};
\fill [red](6.5,7.5) circle (1.5 pt) node[right] {};
\draw[red,-latex](6.98,7.5) -- (8.5,7.5) node[right] {$(x_{N},y_{N}),~\Delta A$};
\fill [red](6.5,0.5) circle (1.5 pt) node[right] {};
\draw[red,-latex](6.98,0.5) -- (8.5,0.5) node[right] {$(x_{N},-y_{N})$};
\fill [orange](4,4) circle (2 pt) node[red,left=5pt,below=2pt] {$O$};
\end{tikzpicture}
\caption{ The detector array consists of  many little detection  elements.}
\end{figure}
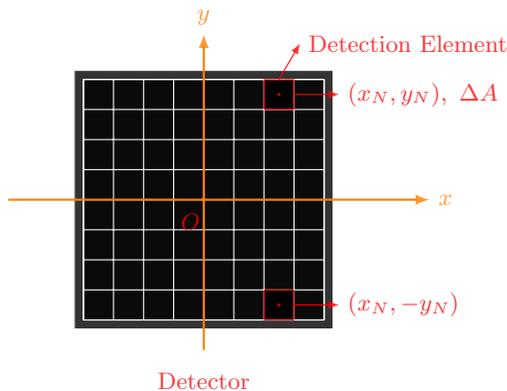

Enlightened by the experiment displaying the different effects of the spin and orbital angular momentum performed by O'Neil et al. \cite{Neil}, we propose here a scheme to examine the different predictions of the canonical and symmetric E-M tensors which is related to the ``local" property of the E-M tensor.  We let the plane of the detector array consist of  many little detection elements.  Each detection element can measure the time-averaged force ${\bm f}_N$ exerted by the light beam on the area of the detection element $\Delta A $ (see FIG. 2).  We represent each detection element with the coordinate $(x_N,y_N) $ of its centre point. The quantity we consider is the displacement of the barycenter of $f^{z}_{N}$, the longitudinal part  of the force ${\bm f}_N$:
\begin{equation}
{\langle y \rangle}_f=\dfrac{\sum_{N} y_N f^{z}_{N}}{\sum_{N} f^{z}_{N}}.
\end{equation}

Theoretically, the force  ${\bm f}_N$  is the time-averaged rate of change of the field momentum received by the  detection element $(x_N,y_N) $. For simplicity, we consider the case that the detector fully absorbs the momentum of light cross the detection plane, so we have
\begin{equation}
f^{z}_{N}=\left\langle\dfrac{{\rm d} P_{N}^{z}(t)}{{\rm d}t}\right \rangle\simeq -\overline{T}^{zz}({x_{N},y_{N}})\Delta A,
\end{equation}
and then the displacement
 \begin{equation}\label{yN}
{\langle y \rangle}_f \simeq \dfrac{\sum_{N}  y\, \overline{T}^{zz}({x_{N},y_{N}})}{\sum_{N} \overline{T}^{zz}({x_{N},y_{N}})}\simeq {\langle y \rangle}_T.
\end{equation}

It is worth noting that the spin angular momentum of light can exert a torque on every detection element along its own symmetric axis, which also produces a  time-averaged force  ${\bm f}_{N}^{\prime}$ on the area $\Delta A$. However, the forces ${\bm f}_{N}^{\prime}$ are identical for the detection elements  $(x_N,y_N) $  and $(x_N,-y_N) $ because of the equivalence of the spin torque on the two detection elements. Therefore,  the forces ${\bm f}_{N}^{\prime}$ do not contribute to the displacement in Eq. (\ref{yN}).

In conclusion, like the geometric spin Hall effect of energy flux, we have demonstrated geometric spin Hall effect of momentum  flux  and proposed a scheme to test this novel effect by the mechanical effect of light. Interestingly, the prediction of the symmetric E-M tensor gives an anomalous effect for the spin-polarized light beam, but that of canonical E-M tensor does not. On the other hand, the orbital angular momentum gives rise to the anomalous GSHEL with respect to momentum flux for both the symmetric E-M tensor and the canonical one. Therefore, for the spin-polarized light beam, we argue that the geometric spin Hall effect with respect to momentum flux can be regarded as an experimental scheme to test the expression of E-M tensor. We strongly urge the experimentalists to perform the measurement we proposed here, not only because it is a novel effect, but also because it contributes to clarifying our understanding of the E-M tensor and angular momentum tensor. 

The authors thank Zi-Wei Chen and De-Tian Yang for helpful discussions. This work is supported by the China NSF via Grants No. 11535005 and No. 11275077.

\end{document}